\documentclass[11pt]{article}
\usepackage{epsfig}
\usepackage{graphicx}

\title{\bf Gluon emission at small longitudinal momenta in the QCD effective action approach}
\author{M.A.Braun, M.I.Vyazovsky\\
Dep. of High Energy physics,
Saint-Petersburg State University,\\
198504 S.Petersburg, Russia}

\newcommand\beq{\begin{equation}}
\newcommand\eeq{\end{equation}}

\newcommand\lra{\leftrightarrow}

\newcommand\ep{\epsilon}

\newcommand\im{{\rm Im}\,}
\def\qa{q_{1+}}
\def\qb{q_{2+}}
\def\ra{r_{1-}}
\def\rb{r_{2-}}

\def\tkb{k_{2\perp}^2}

\def\tp{p_\perp^2}

\newcommand{\Ref}[1]{(\ref{#1})}

\begin{document}

\maketitle
{\bf Abstract}

In the framework of the QCD effective action the vertices of gluon
emission in interaction of reggeons are studied in the limit of small
longitudinal momenta of the emitted gluon. It is found that the vertices
drastically simplify in this limit so that the gluon becomes emitted
from a single reggeon coupled to the projectile and target via
multireggeon vertices. Contribution from this kinematical region is
studied for double and 2$\times$2  elementary collisions inside the
composite projectile and target.

\section{Introduction}
One of the main observables in high-energy collisions of heavy nuclei
is the inclusive cross-section for production of secondaries.
In the perturbative approach it reduces to production of gluons,
which subsequently transform into observed secondary hadrons.
The study of gluon production in the central rapidity region with
transverse momenta much smaller than the longitudinal momenta of the
colliding particles (''Regge kinematics'') has a long history, starting from the
pioneer work on the production of minijets from the BFKL pomeron
~\cite{levin}. Later this problem was studied in the framework of the
dipole picture for the inclusive cross-section in deep-inelastic scattering
(DIS) on the heavy nucleus
~\cite{kovchtuch}, where it was shown that the inclusive cross-section
was related to the so-called unintegrated gluon density in the nucleus.
Still later the formalism of reggeized gluons
(''BFKL-Bartels'' or BFKLB framework ~\cite{bfkl, bfkl1, bartels}) it was demonstrated  that the same cross-section is
consisting of a sum of two
contributions coming from the BFKL pomeron and the cut triple
pomeron vertex ~\cite{braincl}. In collisions of two heavy nuclei
(''AB collisions'') the situation is not so straightforward with the results obtained
only in the framework of the
JIMWLK formalism  (~\cite{jimwlk} and references therein). However, lack of connection with the actual
gluon production in collision of composite targets
and absence of the confirmation in the  BFKLB framework leave certain  open questions
which are waiting for their solution.

In the BFKLB framework gluon production is based on the vertices obtained by the dispersive technique
which uses multiple discontinuities at poles corresponding to intermediate particles. Such vertices
depend only on the transversal momenta of gluons and reggeons. In this form the relation to the scattering on composites is
poorly understood. To describe it one has to rather study  the vertices with the dependence on all 4-momenta included, both transversal and
longitudinal. Such vertices can be found by means of the Lipatov  Effective  Action (LEA) for the QCD ~\cite{lea}, which introduces
the reggeons as independent dynamical variables and describes their interaction with the gluons apart from the standard QCD action.
The simplest vertex for gluon emission from a single reggeon $\Gamma_{R\to RG}$ was constructed in the original BFKL paper~\cite{bfkl,bfkl1}.
In our previous
papers we have found  higher vertices for gluon production in interaction of one and two reggeons $\Gamma_{R\to RRG}$,
$\Gamma_{R\to RRRG}$ and $\Gamma_{RR\to RRG}$. They are quite complicated and derivation of the full inclusive cross-section
in hA and especially AA collisions seems to require extraordinary effort, also taking into account that apart from the contribution from the vertices
one has to calculate numerous contributions from rescattering.

One has to take into account that LEA  only describes interaction  of gluons and reggeons at a given rapidity $y$ (or rather within a finite
interval of rapidities $\Delta y$).  Slices of the total rapidity region separated by large rapidity intervals interact via the exchange of reggeons.
So one has to divide the total rapidity into different number of large intervals connecting different slices of effective action and
obtains different diagrams made of effective vertices and reggeon propagators. Neglecting
the restriction imposed on the width of each slice may lead to divergencies in the integrals over rapidities in the loop integrals.
The practical realization of this picture was first achieved
by the separation from the whole $y$-integration   parts of  small intervals $\Delta y$ in the calculation of the NLO BFKL
kernel in ~\cite{lipfad} where clusters of real gluons were produced in the intermediate state.
Later it was found that in loop calculations  introduction of such direct slicing in rapidity by $\Delta y$ was  inconvenient.
Instead a different method to avoid divergencies due to the limited validity of the LEA vertices was developed based on the
so-called tilted Wilson lines (or tilted light-cone variables). Within this method all different divergencies
coming from the loop integrations in vertices and propagators cancel ~\cite{ref1,ref2,ref3,ref4}.

Vertices studied in this paper refer to a fixed value of rapidity. They do
not contain  internal integrations nor any divergencies. So their calculation can be safely done within the rules of LEA for a given rapidity,
without any  cutoff $\Delta y$ or tilted light-cone technique. We shall find that when the longitudinal momenta of the emitted gluon become small
our vertices acquire a simple  limiting form which actually corresponds to quite different diagrams of the LEA containing the
internal reggeon exchange. These limiting diagrams do not appear in LEA normally and exist only as limits of the standard
contributions. However, their simple structure greatly simplifies the calculation of the contribution to the cross-section in the
limiting domain of momenta of the emitted particles.

The bulk of the paper is devoted to the behavior of the production
vertices at small longitudinal moments (sections 2,3). In section 4 we
study the double inclusive cross-section in the kinematic region where
it is determined by the found degenerate vertices. Section 5 contains
conclusions and some discussion.

\section{Vertices $\Gamma_{R\to RRG}$ and $\Gamma_{R\to RRRG}$ at small $p_-$}
\begin{figure}[h]
\begin{center}
\includegraphics[scale=0.85]{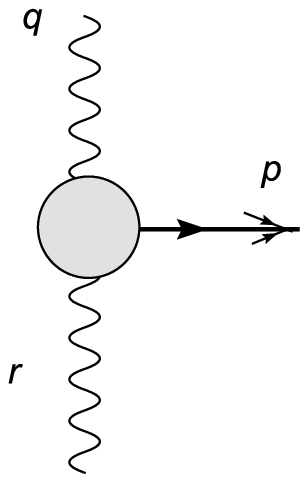}
\end{center}
\caption{Vertex $\Gamma_{R\to RG}$}
\label{incl0}
\end{figure}

To clearly see the problem start with the simplest ("Lipatov") vertex $\Gamma_{R\to RG}$ for the
production of the gluon from reggeon, shown in Fig. \ref{incl0}.
The incoming reggeon with momentum $q$ has $q_+=p_+$ and $q_-=0$, the outgoing reggeon with momentum $r$ has
$r_+=0$ and $r_-=-p_-$. As a result the energy square for the vertex $s_0=(q-r)^2=p^2=0$ and partial energies
$s_1=(q-p)^2$ and $s_2=(p+r)^2$ have orders $p_\perp^2$ considered small relative to the total energy squared $s$ ($|p_\perp|^2/s<<1$).
 This means that the vertex has a finite dimension in rapidity
characterized by the logarithm of these partial energies.

\begin{figure}[t]
\begin{center}
\epsfig{file=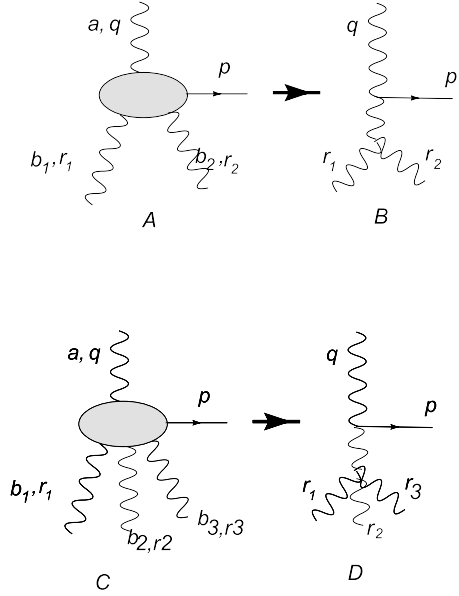, width=6 cm}
\end{center}
\caption{Reduction of the vertices R$\to$RRG and R$\to$RRRG for emission of a particle
(gluon) with a small ''-'' component of its momentum}
\label{fig4a}
\end{figure}

Now consider a more complicated vertex  $\Gamma_{R\to RRG}$ which corresponds to the production
of the gluon when a single reggeon goes into two reggeons.  This vertex enters the amplitude for the gluon production
on a composite state composed of two elementary targets, as we shall see later.
Vertex $\Gamma_{R\to RRG}$  is illustrated in Fig.\ref{fig4a},A.
It was derived in the effective action approach in
~\cite{bravyaz, blsv}:
\beq
\Gamma_{R\to RRG}=W+R+\Big(1\lra 2\Big).
\eeq
Here
\beq
W=-if^{ab_1c}f^{cb_2d}\frac{2q_+q^2}{(q-r_1)^2+i0}B(p,r_2,r_1),
\label{w1}
\eeq
\beq
R=if^{ab_1c}f^{cb_2d}\frac{q^2}{r_{1-}}L(p,r_2).
\label{r1}
\eeq
Notations for momenta  are indicated in Fig.
\ref{fig4a},A; $L$ and $B$ are the Lipatov and Bartels vertices
\beq
L(p,r_2)=\frac{(pe)_\perp}{p_\perp^2}-\frac{(p+r_2,e)_\perp}
{(p+r_2)_\perp^2} , \eeq
\beq
B(p,r_2,r_1)=L(p+r_2,r_1).
\eeq
Here $e$ is the polarization 4-vector with $e_+=0$.
Finally $(1\lra 2)$ means the interchange of outgoing reggeons.

We are going to study the  vertex   in the limiting
region $p_-<<r_{1,2-}$ (region of small $p_-$).  In this case $r_{1-}\simeq -r_{2-}$, so that the "-" momenta of the
two outgoing reggeons
are large and opposite. The '-' momentum comes along one of the outgoing reggeons and goes back along the other.

The part $W$ contains a pole at $(q-r_1)^2\simeq -q_+r_{1-}=0$
that is at $r_{2-}=p_-=\epsilon$. Its contribution to the vertex
contains $\delta(p_--r_{2-})=\delta(p_--\epsilon)$. It is this contribution which
is the only one taken into account in the BFKLB approach, in which discontinuities in energies are
studied. As we move to lower values of $p_-<<r_{2-}$  this $\delta$ function disappears
and only the principal value part of the pole remains.
So at  $p_-<<|r_{1,2-}|$ the denominator
 of (\ref{w1}) simplifies as
\[(q-r_1)^2=(p+r_2)^2=r_2^2+2(pr_2)_{\perp}+2p_+r_{2-}
=r_2^2+2(pr_2)_{\perp}-p_\perp^2\frac{r_{2-}}{p_-} .\]
Since $r_2^2=r_{2\perp}^2$ and $p_\perp$ are assumed to have the same
order of magnitude characteristic to all transverse momenta, at
$p_-<<|r_{2-}|$ one can neglect the first two terms, so that $W$ becomes
\beq
W=if^{ab_1c}f^{cb_2d}\frac{2q_+q^2p_-}{r_{2-}p_\perp^2}B(p,r_2,r_1)=
-i f^{ab_1c}f^{cb_2d}\frac{q^2}{r_{2-}}B(p,r_2,r_1),
\eeq
where we used $q_+=p_+$ and $2p_+p_=-p_\perp^2$. Finally at
 $p_-<<|r_{1,2-}|$ we have $r_{1-}+r_{2-}=0$ and
\beq
W=i f^{ab_1c}f^{cb_2d}\frac{q^2}{r_{1-}}B(p,r_2,r_1).
\eeq
In the sum $W+R$ we find
\beq
W+R=i f^{ab_1c}f^{cb_2d}\frac{q^2}{r_{1-}}
\Big(B(p,r_2,r_1)+L(p,r_2)\Big).
\eeq
We have
\[B(p,r_2,r_1)+L(p,r_2)=\frac{(p+r_2,e)_\perp}{(p+r_2)_\perp^2}-
\frac{(p+r_1+r_2,e)_\perp}{(p+r_1+r_2)_\perp^2}+\frac{(pe)_\perp}
{p_\perp^2}-\frac{(p+r_2,e)_\perp}{(p+r_2)_\perp^2}=L(p,r_1+r_2),\]
so that
\beq
W+R=i f^{ab_1c}f^{cb_2d}\frac{q^2}{r_{1-}}L(p,r_1+r_2) .
\eeq

Adding $(1\lra 2)$ we get
\beq
\Gamma_{R\to RRG}=
i\frac{q^2}{r_{1-}}L(p,r_1+r_2)\Big(
f^{ab_1c}f^{cb_2d}-f^{ab_2c}f^{cb_1d}\Big) ,
\eeq
where we once again used $r_{2-}=-r_{1-}$.
Using the Jacoby identity
\[f^{ab_1c}f^{cb_2d}-
f^{ab_2c}f^{cb_1d}=f^{dac}f^{cb_1b_2}\]
we finally obtain
\beq
\Gamma_{R\to RRG}=
if^{dac}f^{cb_1b_2}\frac{q^2}{r_{1-}}L(p,r_1+r_2).
\label{grrrg}
\eeq
This expression corresponds to the reduction of the vertex  $\Gamma_{R\to RRG}$
as illustrated in the upper part of Fig. \ref{fig4a}.

%%%%%%%%%%%%%%%%%%%%%%%%%%%%%%%%%%%%%%%%%%%%%%%%%%%%%%%%%%%%%%%%%%%%
%%%%%%%%%%%%%%%%%%%%%%%%%%%%%%%%%%%%%%%%%%%%%%%%%%%%%%%%%%%%%%%%%%
The diagram Fig. \ref{fig4a},B by itself does not appear in LEA. The central reggeon in it has both its
two components of the longitudinal momentum equal to zero and so this  diagram lies outside the
standard Regge  kinematics. It only appears as a certain limit of the perfectly legitimate  diagram Fig. \ref{fig4a},A.
One may indicate some arguments to understand its appearance. The initial diagram  Fig. \ref{fig4a},A.
refers to a given rapidity $y$. The coupled reggeons correspond to virtual particles and cannot be characterized by
their rapidities by themselves. However, we can instead consider different partial energies spanned by the interacting
gluon and reggeons: the overall energy $s_0=(q-r_1-r_2)^2=p^2=0$ and also
$s_1=(q-r_1)^2$ and $s_2=(q-r_2)^2$. Then $s_1=-q_+r_{1-}+(q-r_1)_\perp^2$, where we take into account
that $q_-=r_{1,2+}=0$. Since $q_+=p_+=-p_\perp^2/2p_-$, we have
\[s_1=\frac{r_{1-}}{2p_-}p_\perp^2+(q-r_1)_\perp^2 .\]
This energy squared is finite if $p_-$ and $r_{1-}$ have the same order. In this case
the vertex has a finite dimension in energy.
The same is true for the energy squared $s_2$. However, if $p_-\to 0$ energies $s_1$ and $s_2$ become large.
One may think that this is equivalent to a large interval in rapidity covered by the vertex in Fig. \ref{fig4a},A, which
appears outside the allowed region in LEA. According to this logic
in this region one has to divide this interval
in two large ones connected by the reggeon. This will bring us to the diagram with two local vertices shown in Fig. \ref{fig4a},B.
Continuing with this logic one is supposed to introduce a  rapidity cutoff $\Delta y$, insert it somehow into the diagrams and
for small $p_-$   give up the diagram of Fig \ref{fig4a},A and use instead
the diagram with the intermediate reggeon and two vertices of Fig. \ref{fig4a},B.
However, this logic is in fact dubious. There seems to be no clear way to introduce a cutoff $\Delta y$ into the diagram
Fig. \ref{fig4a},A and the diagram Fig. \ref{fig4a},B does not exist in LEA. So the diagram  Fig.\ref{fig4a},A
can be safely used for any values of $p_-$ and Fig. \ref{fig4a},B emerges only as its limit.

%%%%%%%%%%%%%%%%%%%%%%%%%%%%%%%%%%%%%%%%%%%%%%%%%%%%%%%%%%%%%%%%%%%%
%%%%%%%%%%%%%%%%%%%%%%%%%%%%%%%%%%%%%%%%%%%%%%%%%%%%%%%%%%%%%%%%%%%

The same phenomenon occurs with the vertex $\Gamma_{R\to RRRG}$ for
gluon emission in the splitting of a reggeon into three reggeons, which is demonstrated
in detail in Section 3.6 of ~\cite{bpsv0}. In the kinematics $p_-<<r_{1-},r_{2-}, r_{3-}$ it reduces to
\beq
\Gamma_{R\to RRRG}=g^3\Big(f^{adc}f^{cb_1b}f^{bb_3b_2}\frac{1}{r_{1-}(r_{1-}+r_{2-})}
+{\rm permutations\ of}\  r_{1,2}\ {\rm and}\ b_{1,2}\Big) L(p,r_1+r_2+r_3),
\label{rgrrr}
\eeq
which is illustrated in the lower part of Fig. \ref{fig4a}.
%%%%%%%%%%%%%%%%%%%%%%%%%%%%%%%%%%%%%%%%%%%%%%%%%%%%%%%%%%%%%%%%%
%%%%%%%%%%%%%%%%%%%%%%%%%%%%%%%%%%%%%%%%%%%%%%%%%%%%%%%%%%%%%%%%%%%

\section{Vertex RR$\to$ RRG at small $p_\pm$}
The vertex RR$\to$ RRG in the general kinematics was derived in ~\cite{bpsv}.
It is much more complicated than the vertices R$\to$ RRG and R$\to$ RRRG.

We shall demonstrate that at small $p_\pm$ the same reduction holds for this vertex
as for  $\Gamma_{R\to RRG}$ and  $\Gamma_{R\to RRRG}$ discussed in the previous section, which drastically
simplifies the vertices expressing them via the simple BFKL vertex  $\Gamma_{R\to RG}$ and multigluon vertices.
This reduction is illustrated in Fig. \ref{incl48}, above.

\begin{figure}
\begin{center}
\epsfig{file=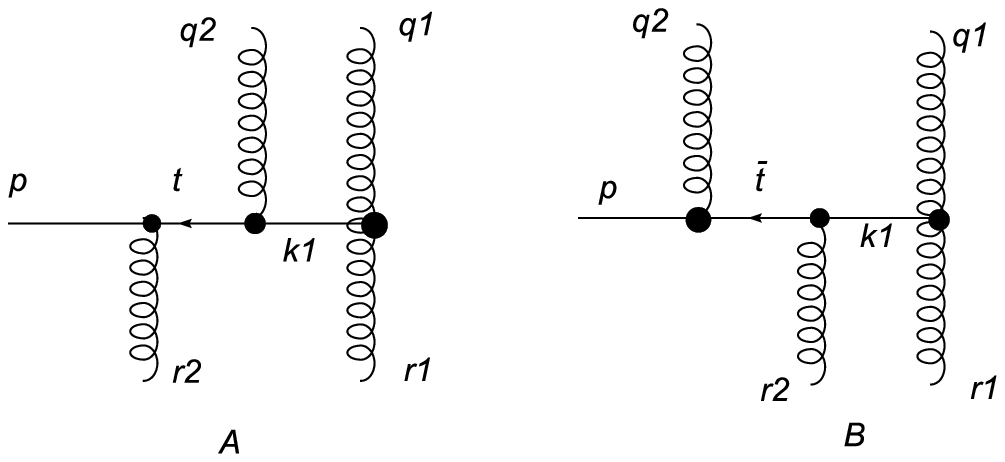, width=8 cm}
\hspace{0.5cm}
%\vspace*  {1 cm}
\epsfig{file=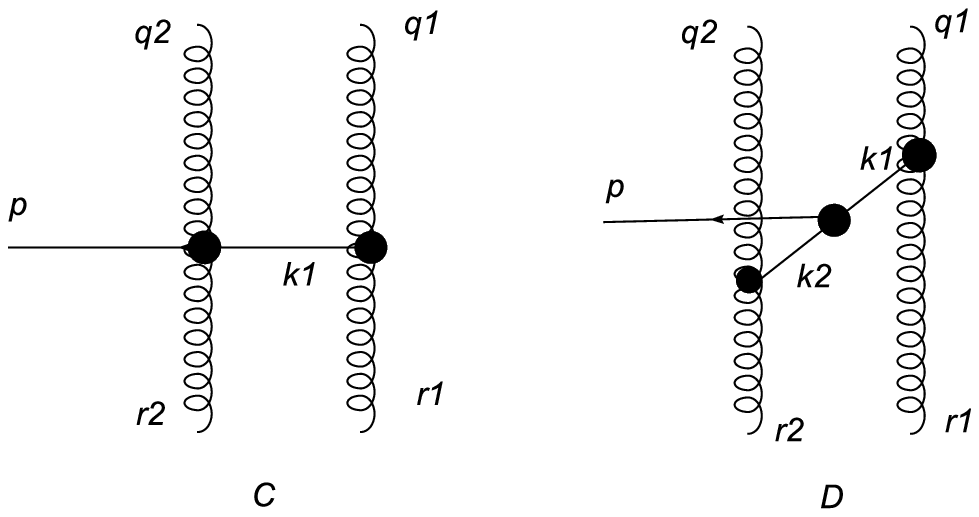, width=8 cm}
\end{center}
\caption{Diagrams for $\Gamma_{RR\to RRG}$}
\label{fig15}
\end{figure}

The vertex RR$\to$ RRG consists of four terms which are shown in Fig. \ref{fig15}.
Explicit expressions for them on the mass shell and multiplied by the polarization vector $\ep$ are given in
~\cite{bpsv}. Below we find expressions for them when
\beq
p_-<<|r_{1-}|,|r_{2-}|,\ \ {\rm and}\ \ p_+<<|q_{1+}|,|q_{2+}| .
\label{kinemab}
\eeq
Some technical details can be found in Appendix.

{\bf 1. Fig. \ref{fig15}, A}
We find from this diagram
\beq
\Gamma_1=-C_1\frac{1}{\qa\ra}(e,q_1+q_2+r_1)_\perp=
-C_1\frac{1}{\qa\ra}(e,p+2r_1+r_2)_\perp.
\label{gamma1}
\eeq
We recall that here $e$ is the polarization 4-vector
with $e_+=0$.

\vspace{1.7cm}
{\bf 2. Fig. \ref{fig15}, B}

From this diagram in the kinematics (\ref{kinemab}) we find
\[
\Gamma_2=C_2\Big\{\frac{1}{2}(pe)_\perp\frac{1}{p_+\ra}-\frac{1}{\qa\ra}
(e,2p-2q_1-q_2)_\perp\]\beq -\frac{1}{\qa\ra}\frac{(pe)_\perp}{\tp}
\Big(4(p,r_1+r_2)_\perp+\bar{t}^2_\perp+2q_1^2-q_2^2+E_0\Big)\Big\}.
\label{gamma2}
\eeq
Here $\bar{t}=p-q_2$.

{\bf 3. Fig. \ref{fig15}, C}

This diagram generates two terms with different color factors.
The corresponding contributions $\Gamma_{3}$
and $\Gamma_{4}$ in the kinematics (\ref{kinemab}) are given by
\beq
\Gamma_3=C_3\Big\{ \frac{1}{p_+\ra}(e p)_\perp+\frac{1}{2}\,\frac{1}{\qa\ra}(e,q_1+r_1)_\perp
+\frac{1}{\qa\ra\tp}(e p)_\perp q_2^2 \Big\}
\label{gamma3}
\eeq
and
\beq
\Gamma_4=C_4\Big\{ -\frac{1}{2}\,\frac{1}{p_+\ra}(ep)_\perp
+\frac{1}{2}\,\frac{1}{\qa\ra}(e,q_1+r_1)_\perp \Big\}.
\label{gamma4}
\eeq

{\bf 4. Fig. \ref{fig15}, D}

In the kinematics (\ref{kinemab}) this diagram gives
\beq
\Gamma_5=C_5\Big\{\frac{1}{2}\,\frac{(pe)_\perp}{p_+}\Big(\frac{1}{\ra}-\frac{1}{\rb}\Big)
-\frac{1}{2\qa\ra}(q_1-r_1-q_2+r_2,e)_\perp\Big\}.
\label{gamma5}
\eeq

The color factors are
\[C_1=f^{db_2c}f^{b_1a_1e}f^{ea_2d},\ \
C_2=f^{da_2c}f^{b_1a_1e}f^{db_2e},\]
\[C_3=f^{a_1b_1d}f^{a_2ce}f^{b_2de},\ \
C_4=f^{a_1b_1d}f^{a_2de}f^{b_2ce},\ \
C_5=f^{a_1d_1b_1}f^{d_1cd_2}f^{a_2d_2b_2} ,\]
so that
\beq
C_3=-C_2,\ \ C_4=C_1
\label{c34}
\eeq
and using the Jacoby identity
\beq
C_5=C_4-C_3=C_1+C_2.
\label{c5}
\eeq

These expressions are to be symmetrized over permutations of reggeons.
Since $\Gamma_1,\dots,\Gamma_4$ are to be symmetrized over all permutations
of momenta and colors of the two incoming and two outgoing reggeons and
 $\Gamma_5$ only over the two outgoing reggeons, we can totally symmetrize
the sum of all diagrams taking the contribution from
 $\Gamma_5$ with weight $1/2$.

First we show that terms of the leading order $1/(p_+ r_{1-})$ and $1/(p_+ r_{2-})$
cancel.
Before symmetrization they are
$$
C_1 \Big[ -\frac{(pe)_{\perp}}{2p_+ r_{1-}}
+\frac{1}{2}\,\frac{(pe)_{\perp}}{2p_+}
\Big( \frac{1}{r_{1-}}-\frac{1}{r_{2-}} \Big) \Big]
+ C_2 \Big[ \frac{(pe)_{\perp}}{2p_+ r_{1-}}
- \frac{(pe)_{\perp}}{p_+ r_{1-}}
+\frac{1}{2}\,\frac{(pe)_{\perp}}{2p_+}
\Big( \frac{1}{r_{1-}}-\frac{1}{r_{2-}} \Big) \Big]
$$
$$
=C_1 \Big[ -\frac{(pe)_{\perp}}{4p_+}
\Big( \frac{1}{r_{1-}}+\frac{1}{r_{2-}} \Big) \Big]
+ C_2 \Big[ -\frac{(pe)_{\perp}}{4p_+}
\Big( \frac{1}{r_{1-}}+\frac{1}{r_{2-}} \Big) \Big]
$$
\begin{equation}
=-\frac{1}{4} \Big( C_1+C_2 \Big)
\Big[ \frac{(pe)_{\perp}}{p_+ r_{1-}}
+ \frac{(pe)_{\perp}}{p_+ r_{2-}} \Big] \ .
\label{es18}
\end{equation}
At all permutations the momentum factor in
\Ref{es18} does not change. So  the leading order contribution
is determined by the symmetrized combinations of the color factors.
One has
\begin{equation}
\mathrm{Sym}\Big\{C_1 + C_2\Big\}=\mathrm{Sym} \Big\{ C_5 \Big\}=
C_5 +C_5(a_1 \lra a_2)
+C_5(b_1 \lra b_2) +C_5(a_1 \lra a_2, b_1 \lra b_2)\ =0.
\label{es13}
\end{equation}
Thus the leading order contribution is indeed canceled.

%%%%%%%%%%%%%%%%%%%%%%%%%%%%%%%%%%%%%%%%%%%%%%%%%%%%%%%%%%%
Next we study the factor multiplying $1/(q_{1+}r_{1-})$. First we separate
terms which do not contain  ${(pe)_{\perp}}/{p^2_{\perp}}$:
$$
-C_1 (e,p +2r_1 +r_2)_{\perp}
-C_2 (e,2p -2q_1 -q_2)_{\perp}
-\frac{1}{2}C_2(e,q_1 +r_1)_{\perp}
+\frac{1}{2}C_1 (e,q_1 +r_1)_{\perp}
$$
$$
-\frac{1}{2}\, \Big( C_1+C_2 \Big)
(e,q_{1}-q_{2}-r_{1}+r_{2})_{\perp}
$$
\begin{equation}
=\Big( -C_1+C_2 \Big) (e,q_{1}+q_{2})_{\perp}
-2C_2 (pe)_{\perp}
+\frac{1}{4} \Big( C_1+C_2 \Big) (pe)_{\perp}\ .
\label{es8}
\end{equation}
Then we transform the factor multiplying the product of
 $1/(q_{1+}r_{1-})$
and ${(pe)_{\perp}}/{p^2_{\perp}}$ (using $\bar t=q_1 -r_1-r_2$):
$$
-C_2\Big(4(p,r_1 + r_2)_{\perp} +\bar t^2_{\perp} +2q_1^2 -q_2^2 +E_0\Big)
-C_2\Big(q_2^2\Big)
$$
\begin{equation}
=-2C_2 (p+r_1+r_2)^2_{\perp} +2C_2 p^2_{\perp}\ .
\label{es9}
\end{equation}
As a result, the total factor multiplying
 $1/(q_{1+}r_{1-})$ takes the form
(using $p+r_1+r_2=q_1+q_2$ and noting that the second term in
 (\ref{es8}) cancels the second term in (\ref{es9}))
\begin{equation}
\Big( -C_1+C_2 \Big) (e,q_{1}+q_{2})_{\perp}
-2C_2\frac{(pe)_{\perp}}{p^2_{\perp}}\,(q_1+q_2)^2_{\perp}
+\frac{1}{4} \Big( C_1+C_2 \Big) (pe)_{\perp}\ .
\label{es10}
\end{equation}

Now we pass to symmetrization. Taking the order of the two
incoming and two outgoing reggeons as  $\Gamma_i\equiv \Gamma_i(2,1|2,1)$ in the expressions
for $\Gamma_i$, $i=1,\dots,5$ we have for the total vertex
\begin{equation}
\sum_{i=1}^{5} \Big[
\Gamma_i(2,1|2,1)+\Gamma_i(1,2|2,1)+\Gamma_i(2,1|1,2)+\Gamma_i(1,2|1,2)
\Big]\ .
\label{esym}
\end{equation}
The transverse factors in  (\ref{es10}) do not change under permutations.
Since in our kinematics with adopted precision $q_{2+}=-q_{1+}$
and $r_{2-}=-r_{1-}$, the denominator  $q_{1+}r_{1-}$ changes sign for the reggeon configurations
$(1,2|2,1)$ or $(2,1|1,2)$ and does not change sign for the configuration $(1,2|1,2)$.
As a result we find antisymmetric combinations of the color factors
\begin{equation}
\mathrm{Asym} \Big\{ C_i \Big\}=
C_i -C_i(a_1 \lra a_2)
-C_i(b_1 \lra b_2) +C_i(a_1 \lra a_2, b_1 \lra b_2)\ .
\label{es11}
\end{equation}

Take first $C_5 =f^{a_1 b_1 d}f^{dce}f^{ea_2 b_2}$
:
$$
\mathrm{Asym}\Big\{ C_5 \Big\} = \mathrm{Asym}\Big\{ C_1 + C_2 \Big\}
$$
\begin{equation}
=f^{a_1 b_1 d}\Big(f^{dce}+f^{ecd}\Big)f^{ea_2 b_2}
-f^{a_2 b_1 d}\Big(f^{dce}+f^{ecd}\Big)f^{ea_1 b_2} =0,
\label{es12}
\end{equation}
from which it follows
$\mathrm{Asym} \Big\{ C_1 \Big\}=-\mathrm{Asym} \Big\{ C_2 \Big\}$.

As a result the last term in \Ref{es10} gives no contribution to the total vertex.
The latter takes the form
\begin{equation}
\frac{\mathrm{Asym} \Big\{ C_2 \Big\}}{q_{1+}r_{1-}}
\Big( 2 (e,q_{1}+q_{2})_{\perp}
-2\frac{(pe)_{\perp}}{p^2_{\perp}}\,(q_1+q_2)^2_{\perp}
\Big)\ .
\label{es14}
\end{equation}

To calculate
\begin{equation}
\mathrm{Asym} \Big\{ C_2 \Big\}=
f^{a_1 b_1 d}f^{db_2 e}f^{ea_2 c}-f^{a_2 b_1 d}f^{db_2 e}f^{ea_1 c}
-f^{a_1 b_2 d}f^{db_1 e}f^{ea_2 c}+f^{a_2 b_2 d}f^{db_1 e}f^{ea_1 c}\ ,
\label{es15}
\end{equation}
one has to apply the Jacoby identity three times.
First consider the difference between the first and third terms.
Applying
$$
f^{a_1 b_1 d}f^{db_2 e}+f^{b_1 b_2 d}f^{da_1 e}+f^{b_2 a_1 d}f^{db_1 e}=0
$$
it can be rewritten as
$-f^{b_1 b_2 d}f^{da_1 e}f^{ea_2 c}$.
Similarly the difference between the fourth an second terms
can be rewritten as
$-f^{b_2 b_1 d}f^{da_2 e}f^{ea_1 c}$. Finally one more Jacoby identity
allows to find
\begin{equation}
\mathrm{Asym} \Big\{ C_2 \Big\}=
f^{b_1 b_2 d} \Big(f^{da_2 e}f^{ea_1 c}-f^{da_1 e}f^{ea_2 c}\Big)
=-f^{a_1 a_2 e} f^{ecd} f^{db_1 b_2}\ .
\label{es16}
\end{equation}

As a result the final result for the total vertex is
\begin{equation}
\Gamma=\frac{f^{a_1 a_2 e} f^{ecd} f^{db_1 b_2}}{q_{1+}r_{1-}}
\Big( -2 (e,q_{1}+q_{2})_{\perp}
+2\frac{(pe)_{\perp}}{p^2_{\perp}}\,(q_1+q_2)^2_{\perp}
\Big),
\label{es17}
\end{equation}
which can be rewritten as
\beq
\Gamma_{RR\to RRG}=2f^{a_1a_2e}f^{ecd}f^{db_1b_2}g^3(q_1+q_2)_\perp^2L(p,r_1+r_2)\frac{1}{q_{1+}r_{1-}}.
\label{rrgrrglim}
\eeq

\section{The inclusive cross-sections in hA and AA collisions at small $p_\pm$}
\subsection{hA collisions}
As we have seen the vertices for gluon production in interaction of reggeons drastically simplify in the region of small
longitudinal momenta of the produced gluon. At first sight it promises to facilitate calculation of physical observables,
such as inclusive  cross-sections  in collision of composite particles, e.g. deuterons. However, we shall discover that
while such facilitation certainly takes place, the resulting cross-sections vanish in this region.

We start with hA collisions.
We recall that the inclusive cross-section from the double scattering of an elementary projectile is  given by the formula
\beq
I_A(p,y)\equiv\frac{2(2\pi)^3d\sigma}{dy d^2pd^2b}=
\frac{A(A-1)}{4\pi k_+s}
\int dz_1dz_2 d\epsilon\cos \Big(\epsilon m(z_1-z_2)/k_+\Big)\im H(p,\epsilon)\rho(b,z_1)\rho(b,z_2),
\label{inclini}
\eeq
where $s=2k_+^2$ is the c.m. energy squared,
$\rho(b,z)$ is the nuclear density, $H(p,\epsilon)$ is the high-energy part of the amplitude left after separating the
nuclear factor, $p$ is the momentum of the emitted particle (gluon) and $\ep$ is the ''-'' component of the
momentum transferred to one of the scattering centers, both in the c.m.system.
The naive Glauber approximation follows if
\beq
{\rm Im}\,H=2\pi \delta(\epsilon)F(p).
\label{h}
\eeq
In which case one gets
\beq
I_A(p,y)=\frac{A(A-1)}{2k_+s}T^2(b)\,F(p).
\label{incl}
\eeq

\begin{figure}
\begin{center}
\epsfig{file=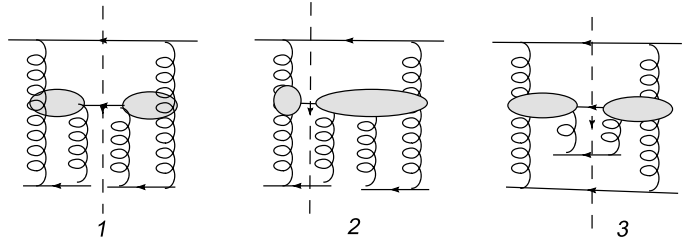, width=9 cm}
\end{center}
\caption{Cut amplitudes with vertices $\Gamma_{R\to RRG}$ and $\Gamma_{R\to RRRG}$}
\label{incl43}
\end{figure}

\begin{figure}
\begin{center}
\epsfig{file=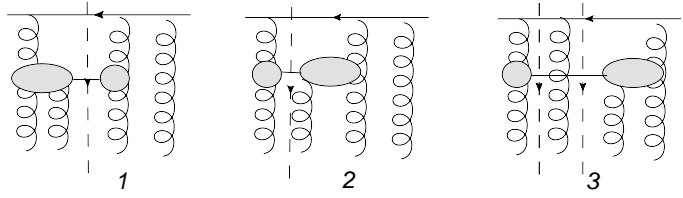, width=9 cm}
\end{center}
\caption{ Typical cut amplitudes with vertex $\Gamma_{R\to RRG}$ and rescattering}
\label{incl44}
\end{figure}

\begin{figure}
\begin{center}
\epsfig{file=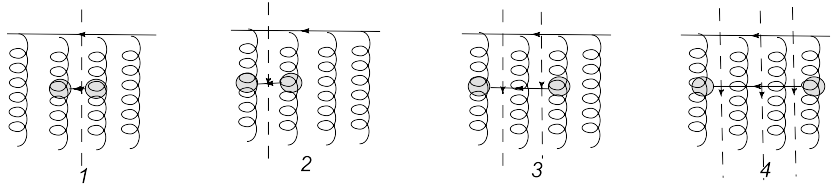, width=11 cm}
\end{center}
\caption{Typical cut amplitudes with pure rescattering}
\label{incl45}
\end{figure}

The high-energy part $H$
can be found
from the scattering amplitude cut to select the observed emitted gluon
in the intermediate state. Due to famous AGK cancelations the gluon can either be emitted from the incoming pomeron or
from the cut triple pomeron vertex. The emission from the pomeron is well-known. Here we are interested
in the emission from the cut triple pomeron vertex which
contains
convolutions of vertices $\Gamma_{R\to RG}$, $\Gamma_{R\to RRG}$ and  $\Gamma_{R\to RRRG}$ shown in Fig. \ref{incl43}.
Apart from this contribution which comes exclusively from the production vertices $\Gamma$ the high-energy part
includes numerous diagrams where one or two reggeons do not interact (''rescattering contribution'') illustrated in
Figs. \ref{incl44} and \ref{incl45}.
In these figures only typical diagrams are shown. We also do not indicate how the outgoing reggeons
are coupled to the two colorless targets. This coupling may be different and follows the pattern of Fig. \ref{incl43}.

\begin{figure}
\begin{center}
\epsfig{file=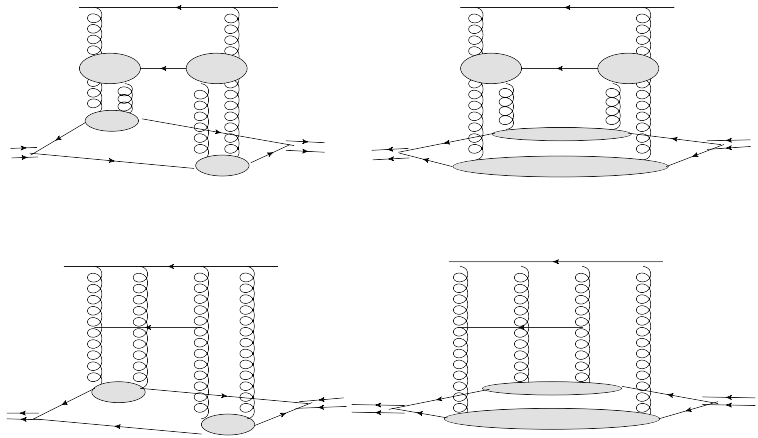, width=10 cm}
\end{center}
\caption{Typical amplitudes with the coupling to the nuclear target explicitly shown }
\label{newincl1}
\end{figure}

We also do not show explicitly evolution of the pomerons attached to the projectile and target (which is
well-known and standardly realized by the BFKL equation)  nor
the actual coupling to colorless scattering centers in the nucleus (in fact nucleons).
One has to take into account that for the heavy nucleus the two centers have to refer to different nucleons.
Otherwise the contribution is down by $A^{1/3}$ assumed large.
For clarity we show some diagrams with the nuclear target explicitly indicated in Fig. \ref{newincl1}.
Several cuts in Figs. \ref{incl44} and \ref{incl45} imply that the sum over contribution of each cut should be taken.
Our aim is to study these contribution in the limit when the longitudinal momentum $p_-$ of the created gluon is much smaller than
the ''-''-momentum $\ep$  transferred to the target, which is the only dimensional variable after the integration over
the longitudinal momenta of the reggeons.

\begin{figure}
\begin{center}
\epsfig{file=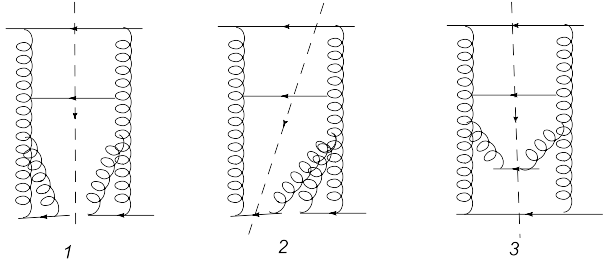, width=8 cm}
\end{center}
\caption{Cut amplitudes with vertices $\Gamma_{R\to RRG}$ and $\Gamma_{R\to RRRG}$
after reduction in Fig. \ref{fig4a}}
\label{crsec1}
\end{figure}

\begin{figure}
\begin{center}
\epsfig{file=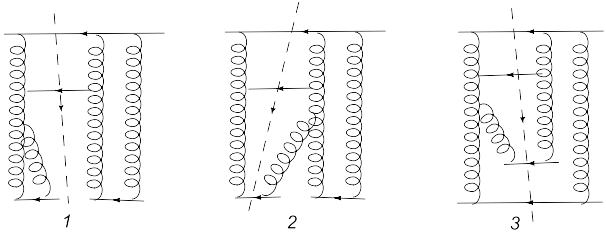, width=8 cm}
\end{center}
\caption{Cut amplitudes with vertex $\Gamma_{R\to RRG}$ and rescattering
after reduction in Fig. \ref{fig4a}}
\label{crsec2}
\end{figure}

As we have shown, in this kinematics
 vertices $\Gamma_{R\to RRG}$ and $\Gamma_{R\to RRRG}$
degenerate into  simple expressions (\ref{grrrg}) and (\ref{rgrrr})
illustrated in Fig. \ref{fig4a}.
The diagrams in Figs. \ref{incl43} and \ref{incl44} then transform into
Figs. \ref{crsec1} and \ref{crsec2}.

Leaving the discussion of rescattering contribution to the end of this section we concentrate here on the diagrams
with the reduced vertices $\Gamma$ in Fig. \ref{crsec1}.
Since
 the pomeron vanishes when the two reggeons are located at
the same point, all diagrams in which the two final reggeons
in $\Gamma_{R\to RRG}$ or $\Gamma_{R\to RRRG}$ are coupled to the same target vanish.
So in the domain $p_-<<|\ep|$ the inclusive cross-section coming from the cut triple pomeron vertex
will be given exclusively by
diagram 3 in Fig. \ref{crsec1}. which corresponds to squaring
vertex $\Gamma_{R\to RRG}$.

Twice imaginary part of the high-energy amplitude $H$ will be given by the square
of the two production amplitudes each containing the vertex $\Gamma_{R\to RRG}$.
It contains  color, energetic,  transverse momentum and numerical
factors.

The color factor is given by (at large $N_c$)
\[ f^{ae_1d}f^{de_2a}f^{e_1b_1b_2}f^{b_1e_2b_2}=
\Big(-N_c\delta_{e_1e_2}\Big)\Big(-N_c\delta_{e_1e_2}\Big)=N_c^4.
\]
Energetic factors come from three cut quark propagators. From the
projectile quark we find $4k_+^2/2k_+=2k_+$. The two target quarks give each
$2l_-$. In the total we get $8k_+l_-^2$. Apart from this we have factors $1/r_{1-}$ in each of the two
vertices $\Gamma_{R\to RRG}$. Taking into account that $r_{1-}$ on the right differ by $\ep$
 we find a longitudinal integral
\beq
J(\ep)=\int\frac{dr_{1-}}{2\pi}
{\cal P}\frac{1}{r_{1-}}\,\cdot\,{\cal P}\frac{1}{r_{1-}-\ep},
\label{intj}
\eeq
where ${\cal P}$ means the principal value.
Calculation gives
\[J(\ep)=
\frac{1}{8\pi}\lim_{\eta_1\to 0,\eta_2\to 0}
\int dx\Big(\frac{1}{x+i\eta_1}+\frac{1}{x-i\eta_1}\Big)
\Big(\frac{1}{x-\ep+i\eta_2}+\frac{1}{x-\ep-i\eta_2}\Big)\]\beq
=-\frac{i}{4}\Big(\frac{1}{\ep-i\eta_1-i\eta_2}
 -\frac{1}{\ep+i\eta_2+i\eta_1}\Big)=\frac{\pi}{2}\delta(\ep)
\label{intjz}
\eeq
and provides factor $(\pi/2)\delta(\ep)$. So actually the diagram
is zero unless $\ep$ is different from zero. However, with $\ep=0$ one cannot realize the
kinematics  $p_-<<|\ep|$, since $p_->0$. This means that in fact the diagram of Fig. \ref{crsec1},3
vanishes in these kinematics.

Later we shall see that the diagrams with rescattering also give no contribution in this domain.

\subsection{AB collisions}

%%%%%%%%%%%%%%%%%%%%%%%%%%%%%%%%%%%%%%%%%%%%%%%%%%%%%%%%%%%%%%%%%%%%%%%%%%%%%%%
%%%%%%%%%%%%%%%%%%%%%%%%%%%%%%%%%%%%%%%%%%%%%%%%%%%%%%%%%%%%%%%%%%%%%%%%%%%%%%%%%%%%%%
%%%%%%%%%%%%%%%%%%%%%%%%%%%%CUT FROM BEFORE%%%%%%%%%%%%%%%%%%%%%%%%%%%%%%%
Passing to AB collisions the obvious generalization of (\ref{incl}) gives the inclusive cross-section in 2$\times$2 collisions
in the Glauber approximation
\beq
I_{AB}(p,b_1,b_2,y)\equiv\frac{(2\pi)^3d\sigma}{dy d^2pd^2b_1d^2b_2}=\frac{A(A-1)B(B-1)}{2k_+2l_-s}FT_A^2(b_1)T_B^2(b_2)
\label{inclab}
\eeq
where $s=2(kl)$, $T_{A,B}(b)$ are the nuclear profile functions and $F$ is
related to the high-energy amplitude $H$ with the ''-'' component $\epsilon$ and ''+'' component $\lambda$ of the transferred
momenta to one of the two target centers and two  projectile centers respectively in the c.m.system by
\beq
{\rm Im}\,H=2\pi \delta(\epsilon)2\pi\delta(\lambda)F+\dots\, .
\label{hab}
\eeq
Terms in $H$ which do not contain $\delta(\epsilon)\delta(\lambda)$ give no contribution
to the inclusive cross-section.

\begin{figure}
\begin{center}
\epsfig{file=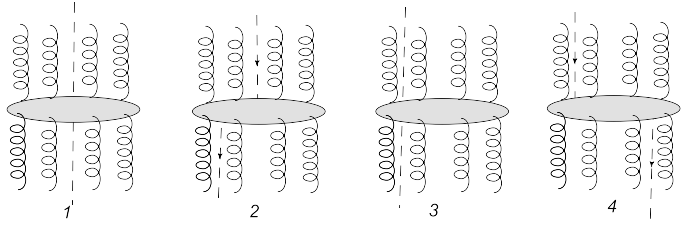, width=9 cm}
\end{center}
\caption{Cut amplitudes with two reggeons attached to each participant}
\label{incl46}
\end{figure}

Due to  AGK cancelations ~\cite{agk} we expect that contributions from emission from the cut pomerons cancels and the final
result comes exclusively from the cut  4-pomeron interaction vertex, including improper terms corresponding to its disconnected
part, corresponding to rescattering in our language. Apart from rescattering and suppressed evolution
each diagram becomes a convolution of two amplitudes for
production of the observed gluon in the transition from 1,2 or 3 initial reggeons into 1,2 or 3 final reggeons.
Then
graphically $H$ corresponds to cut diagrams shown in Fig. \ref{incl46}.
%to which one has to add similar diagrams with interchanges of the participant quarks.
 We do not show the different
ways in which the reggeons may be coupled to the two projectiles and targets. They again follow the pattern illustrated in
Fig. \ref{incl43}.

The diagrams in Fig. \ref{incl46} contain first  ones with a
convolution of two connected vertices for
production of the observed gluon for transition from 1,2 or 3 initial reggeons into 1,2 or 3 final reggeons. Apart from this there
are the diagrams with non-connected vertices, which we interpret as rescattering.
We shall consider the kinematic domain of small longitudinal momenta of the emitted gluon
\beq
p_-<<|\ep|,\ \ p_+<<|\lambda| .
\label{kinemab1}
\eeq
This domain implies small values of $p_\perp^2$ and so refers to emission in the forward  direction.
As we shall argue later  the amplitudes with rescattering do not contribute to the cross-section in the domain
(\ref{kinemab1}).

\begin{figure}
\hspace*{150 pt}
\epsfig{file=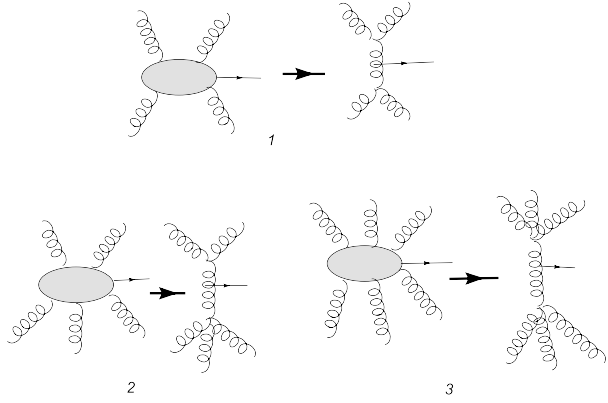, width=8 cm}
\caption{Reduction of the vertices RR$\to$RRG,  RR$\to$RRRG and RRR$\to$ RRRG for emission of a particle
(gluon) with small $\pm$ components of its momentum}
\label{incl48}
\end{figure}

As we have previously shown in the domain (\ref{kinemab}) vertices $\Gamma_{R\to RRG}$, $\Gamma_{R\to RRRG}$ and
$\Gamma_{RR\to RRG}$ degenerate into simple expressions (\ref{grrrg}), (\ref{rgrrr}) and  (\ref{rrgrrglim}) respectively.
Unfortunately  we do not know explicit expressions
for production amplitudes $\Gamma_{RR\to RRRG}$ nor $\Gamma_{RRR\to RRRG}$. However, in the spirit of the QCD effective action
and comparing with cases with smaller numbers of reggeons
we firmly believe  that also for them  a similar reduction takes place. This reduction is graphically shown in the lower part of
Fig. \ref{incl48}.
.

\begin{figure}
\begin{center}
\epsfig{file=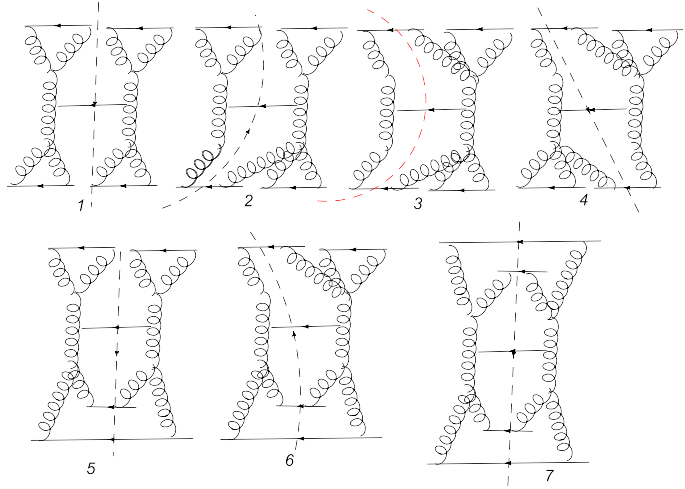, width=9 cm}
\end{center}
\caption{Cut amplitudes for AB scattering with two reggeons attached to each participant
and vertices reduced in the kinematics (\ref{kinemab})}
\label{incl47}
\end{figure}

With thus degenerated vertices, the cut diagrams shown in Fig. \ref{incl46} transform into diagrams illustrated in Fig. \ref{incl47}.
We recall that  when the two reggeons forming a pomeron happen to be located at the same point in the coordinate space the corresponding
pomeron leg vanishes. This excludes all the diagrams in Fig. \ref{incl47} except the last.

As we have seen in the kinematics $p_\pm\to 0$ vertex $\Gamma_{RR\to RRG}$ is simplified to (\ref{rrgrrglim}).
The imaginary part of $H$ will be obtained by squaring this expression together with the relevant energetic, transverse momenta,
numerical and color factors.

Square of the color factor gives $N_c^5$ of which $N_c^4$ are to be included into the coupled pomerons. Energetic factor coming from each projectile quark give $4k_+^2/2k_+=2k_+$
and from each target quark $2l_-$. So we get $16k_+^2l_-^2$ in all. There also remains a
product of $1/q_{1+}r_{1-}$ on the right and on the left, which gives an integral which is a product of integrals
similar to (\ref{intj})
\beq
\int\frac{dq_{1+}dr_{1-}}{4\pi^2}\frac{1}{q_{1+}r_{1-}(\lambda-q_{1+})(\epsilon-r_{1-})}=\frac{\pi^2}{4}\delta(\epsilon)
\delta(\lambda).
\eeq
So we find that
\[
{\rm Im}\,H \propto \delta(\epsilon)\delta(\lambda),\]
which lies outside our kinematics (\ref{kinemab}), since $p_\pm>0$.
So as for $hA$ the domain of small longitudinal moments give nothing for the inclusive cross-section.

\subsection{Rescattering}
The fact  that the diagrams with rescattering do not contribute to the cross-section at small $p_\pm$
follows directly from the number $L$ of longitudinal integrations and dimensional considerations.

Let us start from hA collisions.
All the diagrams with rescattering
have the same order $g^{10}$ as the calculated contribution without rescattering.
Initially the diagram with the total
number $M$ of incoming and outgoing reggeons involves $2(M-3)$ longitudinal momenta of integration, having in mind
that in each participant the sum of transferred momenta is restricted by kinematics.
Each rescattering diminishes  $L$ by 2.
So initially we have $2M-6-2R$ integration variables. In rescattering
each quark propagator gives a $\delta$-function restricting one of the longitudinal momenta.
The total number of these $\delta$-functions is obviously $M-3$. Two more $\delta$-functions
come from the restriction to fix the momentum $p$ of the emitted gluon.
As a result, we find
 \beq
 L=M-2R-5.
 \label{intnum0}
 \eeq
In the main diagram without rescattering
$M=6$, so that $L=1$, which agrees with our previous calculations. The diagrams with a single rescattering
have $M=7$ and $R=1$, which gives $L=0$. Finally with 2 rescatterings
we have $M=8$, $R=2$ and consequently $L=-1$. This negative $L$ means of course that one $\delta$-function survives after
longitudinal integrations. The two remaining longitudinal variables are $p_-$ and $\ep$, so that the result has to be some function
of $p_-$ and $\ep$ of dimension -1.

In the region $p_-<<|\ep|$ the only remaining variable is $\ep$.  So in the case of no integration the result has to be
$\propto 1/\ep$ and in the case of $L=-1$ it has to be $\propto \delta(\ep)$. As we have already mentioned the last case
lies outside the assumed kinematics. If the amplitude is $\propto 1/\ep$ it gives zero in (\ref{inclini}) as it is odd
in $\ep$. Calculations also show that in this case the contribution to the amplitude is real and its imaginary part is zero.
So rescattering amplitudes does not contribute to the inclusive hA cross-section either.

\begin{figure}
\begin{center}
\epsfig{file=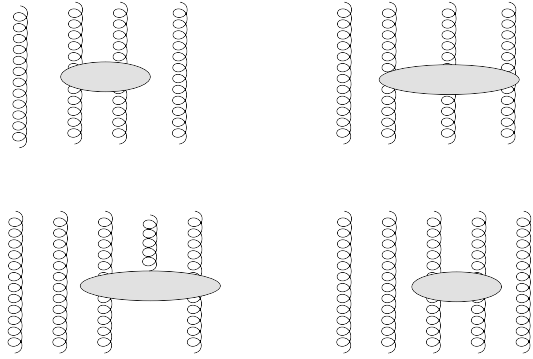, width=7 cm}
\end{center}
\caption{Diagrams with rescattering of orders $g^{10}$ and $g^{12}$}
\label{newincl2}
\end{figure}

This argument works also for AB collisions.
In this case the diagrams with rescattering are many and their order in $g$ starts with $g^{10}$ lower than
for the main diagram calculated in Section 3. In Fig.
\ref{newincl2} we show the diagrams with rescattering of  orders $g^{10}$ and $g^{12}$
lower than order $g^{14}$ considered above.
It is trivial to show that they  all do not give any contribution to the inclusive cross-section
in the domain (\ref{kinemab}) on the same grounds as for hA scattering.

 With 4 participants and $R$ rescatterings
the number of integrations is initially $2(M-4-R)$. Rescattering propagators will produce $M-4$ delta-functions
and together with the restriction imposed by fixing $p$ we get in this case
\beq
L=M-2R-6.
\label{intnum}
\eeq
The order $g^{N}$ of different contributions
may be different  in this case.
It combines the couplings of the reggeons to the participant quarks and the
reggeon interactions, supposed to be only one to correspond to the notion of the
vertex (a single  fixed rapidity $y$). Let the numbers of the incoming and outgoing reggeons be
$M_1$ and $M_2$ respectively With $M=M_1+M_2$ being their total number. $R$ of the reggeons may
go directly from the projectile to the target (rescatterings). The rest of the reggeons,
$M_1-R$ from the projectile and $M_2-R$ from the target interact once with the transitional
kernel $K_{M_1-R\to M_2-R}$, which has order $g^k$ with $k=M_1+M_2-2R-2=M-2R-2$.
Taking into account the coupling to participants $g^M$ we find the total order
\beq
N=2M-2R-2.
\label{order}
\eeq
One, however, has to take into account the number of interacting incoming and outgoing reggeons each
cannot be smaller than 2. This gives a restriction on the number of rescatterings
\beq
R\leq M_1-2,\ M_2-2.
\label{restr}
\eeq

Using  relations (\ref{intnum}) and (\ref{order}) we find the following diagrams with $R$ rescattering having
order $N\leq 14$ interesting for our 4-pomeron vertex. We combine the numbers
$R$, $N$ and $L$ for the diagram into a single index $(R,N,L)$. We find

 {\bf 1.} $M=8:\ \  (0,14,2),\ \ (1,12,0),\ \ (2,10,-2)$;

 {\bf 2.} $M=9:\ \ (1,14,1),\ \ (2,12,-1)$;

 {\bf 3.} $M=10:\ \ (2,14,0),\ \ (3,12,-2)\ ({\rm only}\ M_1=M_2=5)$;

 {\bf 4.} $M=11:\ \ (3,14,-1),\ \ ({\rm only}\ M_1=6,M_2=5)$;

 {\bf 5.} $M=12:\ \ (4,14,-2)\ ({\rm only} M_1=M_2=6)$.

Negative values for $L$ as before imply that one or two $\delta$-functions remain as factors for the diagram.

Inspecting these data we first  find our  main diagram with $M=8$ and no rescattering.
It contains 2  longitudinal integrations as we have previously seen.  We also see that some diagrams with
many rescatterings contain no integrations but one or two $\delta$-functions as factors.
The final dimension of the longitudinal integral is, however, $-2$ in all cases.

In the domain (\ref{kinemab}) the result of the longitudinal integration should  be a function of $\ep$ and $\lambda$
which is Lorenz-invariant and of dimension -2. The only candidates are terms proportional to
$1/\ep\lambda$, $\delta(\ep)/\lambda$, $\delta(\lambda)/\ep$ and $\delta(\ep)\delta(\lambda)$.
In all these cases no contribution to the cross-section follows in the domain (\ref{kinemab}) either because of oddness
in $\ep$ or $\lambda$ or because of violating the kinematics.

\section{Conclusions}
The bulk of our paper is devoted to the study of the vertex for transition RR$\to$RRG in the special
kinematical region where the longitudinal momentum of the emitted gluon is much less than the longitudinal momenta of participating reggeons.
We were able to show that the vertex drastically simplifies, so that emission proceeds from a single intermediate reggeon connected to
the participants via two 3-reggeon vertices.  This also reestablishes the role of the 3-reggeon vertex, absent in many cases due to signature conservation but
appearing under certain kinematical conditions. It also indicates the general rule for gluon emission in interaction of reggeons
when the longitudinal momenta of the emitted gluon turn out to be much smaller (larger) than those of the reggeons. Under this condition
the emission vertex drastically simplifies from a very complicated general expression to a simple and physically clear form.
First examples of such simplification were already mentioned in ~\cite{blsv}. Here we found that it remains valid also for a highly
complicated vertex $\Gamma_{RR\to RRG}$. We firmly believe that this phenomenon holds also for vertices with any number of incoming and
out going reggeons.

Our results have a certain significance in the general theory of
interacting reggeons within the effective action approach. They
 refer to the use of the effective action for calculating not only the vertices at a given rapidity
but also the amplitudes for physical processes at large overall rapidity. Then according to the idea of effective action ont has to divide the total
rapidity interval in slices of definite intermediate rapidities which are then connected by  reggeon exchanges.
It is initially assumed that the vertices determined by the effective action are working only within a given rapidity slice,
so that the set of diagrams describing the amplitude depends on the number of divisions of the total rapidity (resolution in rapidity).
Our results show that this the situation is different: the set of diagrams actually does not depend on the resolution.
Taking the lowest resolution and using the appropriate set of diagrams for the intermediate rapidity one automatically obtains
a different set of diagrams appropriate for higher resolution when one considers the limiting expressions for the initial resolution.
In this sense we prove the independence of the slicing the whole rapidity into partial intervals supposed to be true in the
original derivation of LEA.

Our proof is not complete and does not cover all possible cases. It is based on the vertices which have been explicitly calculated
earlier. In fact the vertices become very complicated with the growth of the number of reggeons and emitted particles.
However, we firmly believe that the result obtained for the considered comparatively simple vertices is true in more complicated cases.

As an application we considered the contribution of the vertices in the limiting cases of higher rapidity resolution to the
calculation of the inclusive cross-section for gluon production. One finds that this contribution is zero. In fact this result
trivially follows from two circumstances. First, in the integration over longitudinal momenta of the reggeons at small $p_\pm$
their order of magnitude automatically reduces to the transferred momenta $\ep$ and $\lambda$. Then condition of
smallness of $p_\pm$ relative to longitudinal momenta of the reggeons transforms into smallness relative to $\ep$ and $\lambda$.
Second, dimensional considerations restrict the final dependence of the longitudinal integrals over "-" components
to either $1/\ep$ or $\delta(\ep)$ and over "+" components to either $1/\lambda$ or $\delta(\lambda)$. Then vanishing of the
contribution in the domain (\ref{kinemab}) immediately follows.

We do not exclude that the obtained properties of the vertices at limiting values of gluon momenta may have other less
trivial applications. We are going to search for such applications in the future study.

%%%%%%%%%%%%%%%%%%%%%%%%%%%%%%%%%%%%%%%%%%%%%%%%%%%%%%%%%%%%%%%%%%%%%%%%%%%
%%%%%%%%%%%%%%%%%%%%%%%%%%%%%%%%%%%%%%%%%%%%%%%%%%%%%%%%%%%%%%%%%%%%%%%%

\section{Appendix. Vertex RR$\to$ RRG in the kinematics (\ref{kinemab})}
\subsection{Contribution from Fig. \ref{fig15}, A}

On mass shell multiplied by the polarization vector $e$ the corresponding amplitude $\Gamma_{1}$ is
given by
\beq
\Gamma_{1}=-C_1\frac{1}{t^2k_1^2}X_1,\ \
X_1=-b\bar{B}-c\bar{C}+d\bar{E},
\label{gam1}
\eeq
The denominator is
\[t^2k_1^2=(-2p_+r_{1-}+t_\perp^2+i0)(-2\qa r_{1-}+k_{1\perp}^2+i0),\]

The coefficients $b,c$ and $e$ are
\[b=2p_+\Big((q_1e)_\perp-(pe)_\perp\frac{\qa}{p_+}\Big)-
2\qa(r_2e)_\perp,\]
\[c=2p_+\Big((q_2e)_\perp-(pe)_\perp\frac{\qb}{p_+}\Big)-
2\qb(r_2e)_\perp,\]
\[d=-2(p+r_2,e)_\perp=-2(te)_\perp.\]
These coefficients do not depend on $\ra$ nor on $\rb$
Terms $\bar{B}$, $\bar{C}$ and $\bar{E}$ are
\[\bar{B}=-4\ra,\ \ \bar{C}=-4\ra+2\frac{r_1^2}{\qa},\]
\[\bar{E}=-2\ra(2\qa+\qb)+q_1^2+q_2^2-k_1^2+r_1^2+(a_1,t+q_2)_\perp
+2r_1^2\frac{\qb}{\qa}-\frac{r_1^2q_2^2}{\qa\ra},\]
where $a_1=q_1+r_1$.

We rewrite
\[X_1=-(b+c)\bar{B}-2c\frac{r_1^2}{\qa}+d\bar{E}\equiv X_{11}+X_{12}+X_{13}.\]
So
\[X_{11}=-(b+c)\bar{B}=8\ra p_+(r_1e)_\perp.\]
with
$
b+c=2p_+(r_1e)_\perp$
Next we find
\[X_{12}=-2c\frac{r_1^2}{\qa}=-4\frac{r_1^2}{\qa}
\Big(p_+(q_2e)_\perp-\qb(te)_\perp\Big)=
4p_+(k_1e)_\perp\frac{r_1^2}{\qa}-4r_1^2(te)_\perp
\]
and finally
\[X_{13}=2(te)_\perp\Big(2\ra(2\qa+\qb)-q_1^2-q_2^2+r_1^2-k_1^2-2p_+\frac{r_1^2}{\qa}-(a_1,t+q_2)_\perp
+\frac{q_2^2r_1^2}{\qa\ra}\Big).\]

The leading terms in our kinematics are proportional to $1/\qa\ra$. So terms in $X_1$ which grow slowlier
than $p_+\ra$ can be dropped. In particular term $X_{12}$ can be dropped altogether. Separating the
longitudinal terms in $\bar{E}$ we have
\[\bar{E}=-2p_+\ra+q_1^2+q_2^2-k_{1\perp}^2-r_1^2+(a_1,t+q_2)_\perp+
2r_1^2\frac{p_+}{\qa}-\frac{r_1^2q_2^2}{\qa\ra}.
\]
Only the first term contributes after multiplication by $d$.

Combining all terms we have
\[X_1=4p_+\ra(e,q_1+q_2+r_1)_\perp,\]
which gives the final result (\ref{gamma1}).

\subsection{Contribution  from Fig. \ref{fig15}, B}

On mass shell multiplied by the polarization vector $e$ the corresponding amplitude $\Gamma_{2}$ is
given by
\beq
\Gamma_2=-C_2\frac{1}{\bar{t}^2k_1^2}X_2,\ \ X_2=\bar{a} A+\bar{b} B+\bar{c} C+\bar{d} E.
\eeq

The denominator is
\[\bar{t}^2k_1^2=(2\qa p_-+\bar{t}_\perp^2)(-2\qa r_{1-}+k_{1\perp}^2+i0).\]
The coefficients $\bar{a},...\bar{e}$ are
\[\bar{a}=(pe)_\perp\frac{\bar{t}^2}{p_+}\ \  {\rm where}\ \
\bar{t}^2=2\qa p_-+\bar{t}_\perp^2,\]
\[\bar{b}=2p_-(r_1e)_\perp+2\ra\Big((q_2e)_\perp-
(pe)_\perp\frac{\qb}{p_+}\Big)+2(pe)_\perp
\Big(\ra-\ra\frac{q_2^2}{p_\perp^2}+\frac{(pr_1)_\perp}{p_+}\Big),\]
\[\bar{c}=2p_-(r_2e)_\perp+2\rb\Big((q_2e)_\perp-
(pe)_\perp\frac{\qb}{p_+}\Big)+2(pe)_\perp
\Big(\rb-\rb\frac{q_2^2}{p_\perp^2}+\frac{(pr_2)_\perp}{p_+}\Big),\]
\[\bar{d}=2(q_2e)_\perp+2(pe)_\perp\Big(1-\frac{\qb}{p_+}-
\frac{q_2^2}{p_\perp^2}\Big).\]
Furthermore
\[ A=3\qa-\frac{q_1^2}{\ra},\ \ B=4\qa,\ \ C=4\qa-2\frac{q_1^2}{\ra}\]
\[E
=-2\qa(2\ra+\rb)+r_2^2+r_1^2-k_1^2+q_1^2+2q_1^2\frac{\rb}{\ra}-
(a_1,\bar{t}-r_2)_\perp-\frac{q_1^2r_2^2}{\ra\qa}.
\]

We present
\[X_2=\bar{a}A+(\bar{b}+\bar{c})B-2\bar{c}\frac{q_1^2}{\ra}+\bar{d}E\equiv
X_{21}+X_{22}+X_{23}+X_{24}.
\]

In our limit we can drop the second term in $\bar{a}$. So
\[
X_{21}=3(pe)_\perp\frac{\qa}{p_+}\bar{t}^2=
6(pe)\frac{\qa^2p_-}{p_+}+3(pe)\frac{\qa}{p_+}\bar{t}^2_\perp .
\]

Next
we find
\[\bar{b}+\bar{c}=2p_-\Big[(e q_1)_\perp-(e p)_\perp\frac{\qa}{p_+}-
(e p)_\perp\Big(1-\frac{q_2^2}{\tp}+2\frac{(p,r_1+r_2)_\perp}{\tp}\Big)\Big],
\]
so that
\[
X_{22}=-8(pe)_\perp\frac{\qa^2p_-}{p_+}+
8\qa p_-\Big[ (e q_1)_\perp-(e p)_\perp\Big(1-\frac{q_2^2}{\tp}+2\frac{(p,r_1+r_2)_\perp}{\tp}\Big)\Big].
\]

Next we find the terms of interest in part $X_{23}$
\[
X_{23}=4(pe)_\perp\frac{\qa}{p_+}q_1^2.
\]

Finally the most complicated term $X_{24}$. Separating the longitudinal momenta in $E$ we
find terms which do not vanish at $\qa,\ra\to\infty$
\[
E=2p_-\qa+E_0,\ \ E_0=r_1^2+r_2^2-q_1^2-k^2_{1\perp}-(q_1+r_1,\bar{t}-r_2)_\perp
\]
and multiplying by $\bar{d}$ we find
\[
X_{24}=4(pe)_\perp\frac{\qa^2p_-}{p_+}+2(pe)_\perp\frac{\qa}{p_+}E_0+
4\qa p_-\Big((q_2e)_\perp-(pe)_\perp\frac{q_2^2}{\tp}\Big).
\]

Summing all terms we have
\[ X_2=2(pe)_\perp\frac{\qa^2p_-}{p_+}+
(pe)_\perp\frac{\qa}{p_+}\Big(3\bar{t}^2_\perp+4\tp-2q_2^2+4q_1^2+8(p,r_1+r_2)_\perp+2E_0\Big)
+4\qa p_-(2q_1+q_2,e).
\]
%{\it (Checked numerically)}

We have to
take into account that the denominator $\bar{t}^2$ has to be taken with the next order correction
\[\frac{1}{\bar{t}^2}=\frac{1}{2\qa p_-+\bar{t}^2_\perp}=\frac{1}{2\qa p_-}\Big(1-\frac{\bar{t}^2_\perp}{2\qa p_-}\Big).\]
As a result
\[
\frac{X_2}{\bar{t}^2k_1^2}=-\frac{1}{4\qa^2\ra p_-}\Big(X_2-(pe)\frac{\qa}{p_+}\bar{t}^2_\perp\Big),\]
so that we get
\beq
\Gamma_2=-C_2\frac{\tilde{X}_2}{4\qa^2\ra p_-}
\eeq
where
\beq
\tilde{X}_2=2(pe)_\perp
+(pe)_\perp
\frac{\qa}{p_+}\Big(2\bar{t}^2_\perp+4\tp-2q_2^2+4q_1^2+8(p,r_1+r_2)_\perp+2E_0\Big)
+4\qa p_-(2q_1+q_2,e).
\eeq
%{\it (Checked numerically)}

$\Gamma_2$ can be rewritten in the form
\[
\Gamma_2=C_2\Big\{\frac{1}{2}(pe)_\perp\frac{1}{p_+\ra}-\frac{1}{\qa\ra}
(e,2p-2q_1-q_2)_\perp\]\beq -\frac{1}{\qa\ra}\frac{(pe)_\perp}{\tp}
\Big(4(p,r_1+r_2)_\perp+\bar{t}^2_\perp+2q_1^2-q_2^2+E_0\Big)\Big\}.
\eeq
Here $\bar{t}=p-q_2$.

\subsection{Contribution from Fig. \ref{fig15}, C}

This diagram generates two terms with different color factors.
The corresponding amplitudes $\Gamma_{3}$
and $\Gamma_{4}$ are given by
given by
\beq
\Gamma_{3,4}=C_{3,4}\frac{1}{k_1^2}X_{3,4},
\eeq
where  the denominator is
$k_1^2=-2\qa r_{1-}+k_{1\perp}^2+i0$.

We have
\[
X_3=-(a_1e)_\perp
+2\frac{(pe)_\perp}{p_+}\Big(\frac{q_1^2}{r_{1-}}-q_{1+}\Big)-
2\frac{(pe)_\perp q_2^2}{p_\perp^2\rb}\Big(\frac{r_1^2}{q_{1+}}
-r_{1-}\Big)\]
and
\[
X_4=-(a_1e)_\perp
-\frac{(pe)_\perp}{p_+}\Big(\frac{q_1^2}{r_{1-}}-q_{1+}\Big)
+\frac{(pe)_\perp q_2^2}{p_+\ra\rb}\Big(\frac{r_1^2}{q_{1+}}
-r_{1-}\Big).\]
Here $a_1=q_1+r_1$.

All terms in $X_{3,4}$ which are small in the limit $\qa,\ra \to\infty$ can be dropped.
Then we obtain in the straightforward manner (\ref{gamma3}) and (\ref{gamma4}).

\subsection{Contribution from Fig. \ref{fig15}, D}

On mass shell and convoluted with the polarization vector the corresponding amplitude $\Gamma_{5}$
is given by
\beq
\Gamma_{5}=C_5\frac{1}{k_1^2k_2^2}X_5,\ \
X_5=2(k_2L_1)L_{2}-
2(k_1L_2)L_{1}+(L_1L_2)(k_1-k_2)_e \, .
\label{a5}
\eeq
Here $k_{1,2}=q_{1,2}-r_{1,2}$,  $C_5=C_1+C_2$.
The denominator is
\[
k_2^2k_1^2=(-2\qb \rb+\tkb+i0)(-2\qa r_{1-}+k_{1\perp}^2+i0).
\]
The Lipatov vertices convoluted with polarization vectors are
\[L_1=(a_1 e)_\perp-\frac{(pe)_\perp}{p_+}\Big(\frac{q_1^2}{\ra}-\qa\Big),
\ \ L_2=(a_2 e)_\perp-\frac{(pe)_\perp}{p_+}\Big(\frac{q_2^2}{\rb}-\qb\Big),\]
\[(k_1-k_2)_e=(k_1-k_2,e)_\perp-\frac{(pe)_\perp}{p_+}(\qa-\qb),\]
where $a_1=q_1+r_1$ and $a_2=q_2+r_2$.

One  finds
\beq (k_2L_1)=(pL_1)=-p_+\ra-p_-\qa+(pa_1)_\perp+r_1^2\frac{p_+}{\qa}+
q_1^2\frac{p_-}{\ra},
\label{eq81}
\eeq
\beq (k_1L_2)=(pL_2)=-p_+\rb-p_-\qb+(pa_2)_\perp+r_2^2\frac{p_+}{\qb}+
q_2^2\frac{p_-}{\rb}
\label{eq81a}
\eeq
and finally
\[(L_1L_2)=(a_1a_2)_\perp+\qa\rb+\qb\ra-r_1^2\frac{\qb}{\qa}
-r_2^2\frac{\qa}{\qb}-q_1^2\frac{\rb}{\ra}-q_2^2\frac{\ra}{\rb}
+\frac{q_1^2r_2^2}{\ra\qb}+\frac{q_2^2r_1^2}{\rb\qa}.
\]

In our limit we get
\[
L_2(k_2L_1)=(pe)_\perp\frac{\qb}{p_+}(-p_+\ra-p_-\qa)+
(pe)_\perp\frac{\qb}{p_+}(pa_1)_\perp+(a_2e)_\perp(-p_+\ra-p_-\qa)
\]
and
\[
L_1(k_1L_2)=(pe)_\perp\frac{\qa}{p_+}(-p_+\rb-p_-\qb)+
(pe)_\perp\frac{\qa}{p_+}(pa_2)_\perp+(a_1e)_\perp(-p_+\rb-p_-\qb).
\]
Taking the difference we find
\[L_2(k_2L_1)-L_1(k_1L_2)=
(pe)_\perp(\qa\rb-\qb\ra)+(pe)_\perp\frac{1}{p_+}
\Big(\qb(pa_1)_\perp-\qa(pa_2)_\perp\Big)\]\[+
(a_2e)_\perp(-p_+\ra-p_-\qa)-(a_1e)_\perp(-p_+\rb-p_-\qb).
\]
Only terms quadratic in $\qa=-\qb$ and $\ra=-\rb$ can give  a non-zero contribution in our limit.
However, as we see, they are canceled in the first term. So we do not get any contribution from
the first two terms in $X_5$

In the third term we can leave only the leading term in $(L_1L_2)$
\[
(L_1L_2)=\qa\rb+\qb\ra,
\]
which leads to
\[X_5=-\frac{(pe)_\perp}{p_+}(\qa-\qb)(\qa\rb+\qb\ra)
+(k_1-k_2,e)_\perp(\qa\rb+\qb\ra).\]

In the second term we can take $\qb=-\qa$ and $\rb=-\ra$ to finally get
\beq
\Gamma_5=C_5\Big\{\frac{1}{2}\,\frac{(pe)_\perp}{p_+}\Big(\frac{1}{\ra}-\frac{1}{\rb}\Big)
-\frac{1}{2\qa\ra}(q_1-r_1-q_2+r_2,e)_\perp\Big\}.
\eeq
Note that the first term is the only one
where the difference between $r_{1-}$ and $-r_{2-}$ is significant.

%%%%%%%%%%%%%%%%%%%%%%%%%%%%%%%%%%%%%%%%%%%%%


\begin{thebibliography}{99}

\bibitem{levin} L.V.Gribov, E.M.Levin, M.G.Ryskin, Phys. Lett. {\bf B 121} (1983) 65.
\bibitem{kovchtuch} Yu.V.Kovchegov, K.Tuchin, Phys. Rev. {\bf D 65} (2002) 074026.
\bibitem{bfkl} E.A.Kuraev, L.N.Lipatov, V.S.Fadin, Sov. Phys. JETP {\bf 45} (1977) 199.
\bibitem{bfkl1} I.I.Balitski, L.N.Lipatov, Sov. J. Nucl. Phys. {\bf 28} (1978) 822.
\bibitem{bartels} J.Bartels, Nucl. Phys. {\bf B 151} (1979) 293.
\bibitem{braincl} M.A.Braun, Eur. Phys. J. {\bf C 48} (2006) 501.
%\bibitem{running1} W.A.Horowitz, Yu.V.Kovchegov, Nucl. Phys. {\bf A 849} (2011) 72.
%\bibitem{running2} M.A.Braun, Eur. Phys. J. {\bf C 75} (2015) :298.
\bibitem{jimwlk} K.Dusling, F.Gelis, T.Lappi, R.Venugopalan, Nucl. Phys. {\bf A 836} (2010) 159.
\bibitem{lea} L.N.Lipatov, Phys. Rep. {\bf 286} (1997) 1997.
\bibitem{lipfad} V.S.Fadin, L.N.Lipatov, Phys. Lett. {\bf B 429} (1998) 127.
\bibitem{ref1} M.Hentschinski, A.Sabio Vera, Phys. Rev. {\bf D 85} (2012) 056006;
arXiv:1110.6741 [hep-th].
\bibitem{ref2} G.Chachamis, M.Hentschinski, J.D.Madrigal Martinez, A.Sabio Vera, \\
Phys.Rev. {\bf D 87} (2013) 076009; arXiv:1212.4992 [hep-th].
\bibitem{ref3} G.Chachamis, M.Hentschinski, J.D.Madrigal Martinez, A.Sabio Vera, \\
Nucl.Phys. {\bf B 876} (2013) 453; arXiv:1307.2591 [hep-th].
\bibitem{ref4} M.Nefedov, V.Saleev, Mod. Phys. Lett {\bf A 32} (2017) 1750207;
arXiv:1709.06246 [hep-th].
%\bibitem{bartels1} J.Bartels, Nucl. Phys. {\bf B 175} (1980) 365.
%\bibitem{barwue} J.Bartels, M.Wuesthoff, Z.Physik {\bf C 66} (1995) 157.
%\bibitem{mueller} A.Mueller, B.Patel, Nucl. Phys. {\bf B 425} (1994) 471.
%\bibitem{bravac} M.A.Braun, G.P.Vacca, Eur. Phys. J. {\bf C 6} (1999) 147.
%\bibitem{barew} J.Bartels, C.Ewerz, JHEP {\bf 9909} (1999) 926.
%\bibitem{brauntmf} M.A.Braun, Theor. and Math. Phys. {\bf 148} (2006) 923.
\bibitem{agk} V.A.Abramovsky, V.N.Gribov, O.V.Kancheli,
 Sov. J. Nucl. Phys. {\bf 18}(1974) 308.
\bibitem{bravyaz} M.A.Braun, M.I.Vyazovsky, Eur. Phys. J. {\bf C 51} (2007) 103.
\bibitem{blsv} M.A.Braun, L.N.Lipatov, M.Yu.Salykin, M.I.Vyazovsky,
 Eur. Phys. J. {\bf C 71} (2011) :1639.
%\bibitem{bsv} M.A.Braun, M.Yu.Salykin, M.I.Vyazovsky, Eur. Phys. J. {\bf C 72} (2012) :1864.
\bibitem{bpsv0} M.A.Braun, S.S.Pozdnyakov, M.Yu.Salykin, M.I.Vyazovsky,
 Eur. Phys. J. {\bf C 72} (2012) :2223.
%\bibitem{bal} I.Balitski, Nucl. Phys. {\bf B 463} (1996) 99.
%\bibitem{kov} Yu.V. Kovchegov, Phys. Rev. {\bf D 60} (1999) 034008.
%\bibitem{brauncalc} M.A.Braun,  Eur. Phys. J. {\bf C 39} (2005) 451.
\bibitem{bpsv} M.A.Braun, S.S.Pozdnyakov, M.Yu.Salykin, M.I.Vyazovsky,
 Eur. Phys. J. {\bf C 75} (2015) :222.
%\bibitem{barrys} J.Bartels, M.G.Ryskin, Z.Phys. {\bf C 76} (1997) 241.
\end{thebibliography}
\end{document}